\documentclass[reprint,amsmath,amssymb,aps,pra]{revtex4-2}
\usepackage[colorlinks,linkcolor=blue,anchorcolor=blue,citecolor=blue]{hyperref}
\usepackage{graphicx}
\usepackage{amsmath}
\usepackage{braket}

\begin{document}
\title{Nonreciprocal Landau-Zener Tunneling in the Presence of Heat-Bath-Induced Colored Noise}
\author{Yi Cao, Jie Liu}
\email[]{jliu@gscaep.ac.cn}
\affiliation{Graduate School, China Academy of Engineering Physics, Beijing 100193, China}
\date{\today}

\begin{abstract}
In this work, we investigate the effects of heat-bath-induced colored noise on nonreciprocal Landau-Zener tunneling. We perform numerical calculations of the instantaneous energy levels and find that noise induces stochastic fluctuations in these levels. Beyond such fluctuations, sufficiently large nonreciprocity parameters enable noise to further trigger the emergence of multiple exceptional points. The corresponding tunneling probability is significantly modified by colored noise. On the one hand, colored noise generally enhances the tunneling probability under nonadiabatic sweeping. On the other hand, it can break the reciprocity of Landau-Zener tunneling in the adiabatic limit. In particular, for large nonreciprocity parameters, the adiabatic tunneling probability is insensitive to noise and determined by exceptional points. We plot the phase diagram of tunneling probability as a function of the sweep rate and nonreciprocity parameter, for different amplitudes and correlation times of colored noise. Analytically, we derive the exact expression for the tunneling probability in the absence of noise and an approximate solution in the white-noise limit, employing Weber functions and the Wiener-Hermite expansion method. Some implications of our theory are discussed.
\end{abstract}

\maketitle

\section{Introduction}
The nonreciprocal Landau-Zener (LZ) model, as an important extension of the standard LZ framework, generalizes the system to the non-Hermitian regime via nonreciprocal energy-level coupling~\cite{Morales-Molina2011, Torosov2017, Shen2019, Wang2022, Tong2023, Dong2023, Wang2023, Cao2024, Chang2025, Cheng2026}. In this non-Hermitian system, exceptional points (EPs) emerge where two eigenvalues and their corresponding eigenvectors coalesce~\cite{Konotop2016, Ashida2020, Bergholtz2021}. This exotic degeneracy gives rise to some intriguing phenomena: even in the adiabatic limit where the energy-level bias varies extremely slowly with time, the evolution of a quantum state can fail to follow the corresponding eigenstate~\cite{Wang2022, Wang2023}; in driven non-Hermitian photonic Lieb lattices, higher-order EPs can lead to asymmetric energy transitions, pseudo-Hermitian propagation and secondary emission~\cite{Xia2021}, to name only a few. On the other hand, the nonreciprocal coupling has been experimentally realized in optical systems~\cite{Guo2009, Ruter2010, Lira2012, Regensburger2012, Xu2016}, superconducting quantum circuits~\cite{Sliwa2015, Lecocq2017}, cold atomic systems~\cite{Sayrin2015, Li2023} and other related physical platforms.

In realistic physical environments, quantum systems are unavoidably coupled to a heat bath. The resulting thermal noise induces dissipation and decoherence~\cite{Gardiner2004, Breuer2007, Weiss2021}, thereby altering quantum dynamical evolution. Such noise-induced dynamical processes have been investigated within the standard LZ framework~\cite{Grifoni1998, Ivakhnenko2023, Kayanuma1984a, Kayanuma1984b, Pokrovsky2003, Kenmoe2013}. On the other hand, the nonreciprocal systems whose non-Hermitian dynamics are characterized by the EPs, are expected to exhibit entirely distinct responses to noise. In the vicinity of an EP, these systems exhibit power-law-enhanced sensitivity to parameter perturbations~\cite{Ganainy2018, Bergholtz2021}, which significantly amplifies the influence of environmental noise~\cite{Miri2019}. Nevertheless, when the system traverses the EPs at a finite rate, how noise affects the evolution of quantum states remains elusive. 

In the present work, we investigate the nonreciprocal LZ tunneling dynamics under bath-induced colored noise, to address how the interplay between the non-Hermiticity and environmental noise affects the quantum tunneling dynamics. We perform numerical calculations of the instantaneous energy levels and calculate the tunneling probability as a function of the sweep rate and nonreciprocity parameter, for different amplitudes and correlation times of colored noise. We find that the nonreciprocal quantum tunneling is significantly modified by colored noise. We further adopt Weber functions and the Wiener-Hermite expansion method to derive the analytical expressions for the tunneling probability in the noiseless case and the white-noise limit, respectively. This paper is organized as follows. Section~\ref{Sec-Model} establishes the theoretical model and defines the tunneling probability for different sweep directions. Section~\ref{Sec-Numerical Results} presents the numerical results, and Sec.~\ref{Sec-Analytical Results} offers relevant analytical derivations. A brief summary is provided in Sec.~\ref{Sec-Conclusion}. Appendix~\ref{Appendix-1} and \ref{Appendix-2} detail the derivation of the stochastic Schr\"odinger equation and the approximate tunneling probability in the white-noise limit, respectively. 

\section{Model}\label{Sec-Model}
In this work, we consider a nonreciprocal LZ model coupled to a heat bath. The total Hamiltonian is given by
\begin{equation}
  H=H_{S}+H_{B}+H_{I},
\end{equation}
where the system Hamiltonian $H_{S}$ takes the form~\cite{Morales-Molina2011, Torosov2017, Wang2022}
\begin{equation}\label{Eq-System Hamiltonian}
  H_{S}=-\frac{1}{2}\begin{pmatrix} \alpha t & v \\ v(1-\delta) & -\alpha t \end{pmatrix}.
\end{equation}
Here, the level bias is assumed to vary linearly in time at a constant rate $\alpha$. $v$ represents the coupling strength between the two energy levels, $\delta>0$ is the nonreciprocity parameter that results in non-Hermiticity. The Hamiltonian admits a rescaling by $v$, allowing us to normalize the energy unit to $v=1$ without loss of generality. To simplify the analysis, we model the bath as a collection of harmonic oscillators with Hamiltonian $H_{B}=\sum_{k}\left(\frac{p_{k}^{2}}{2}+\frac{1}{2}\omega_{k}^{2}q_{k}^{2}\right)$~\cite{Ford1965, Gardiner2004}, where $q_{k}$ and $p_{k}$ are position and momentum operators of the $k$-th bath oscillator, and $\omega_{k}$ is the corresponding angular frequency. The system-bath interaction Hamiltonian $H_{I}$, which characterizes the linear coupling arising from induced shifts in the equilibrium positions of the bath oscillators due to state transitions of the two-level system~\cite{Kayanuma1984a, Leggett1987}, can be written as $H_{I}=-\frac{1}{2}\sum_{k}\eta_{k}q_{k}\sigma_{z}$, where $\eta_{k}$ is the coupling constant, and $\sigma_{z}$ is the Pauli matrix.

Within the Markovian approximation, the bath is treated in the classical limit $\hbar\to0$ and modeled by the Drude spectral density $\eta\omega\gamma/(\omega^{2}+\gamma^{2})$~\cite{Ford1988, Breuer2007, Weiss2021}, where $\eta$ characterizes the effective system-bath coupling strength and $\gamma$ denotes the Drude cutoff frequency. The resulting effective stochastic Schr\"odinger equation governing the system dynamics takes the following form (here we set $\hbar=1$), with detailed derivations given in Appendix~\ref{Appendix-1},
\begin{equation}\label{Eq-SSE}
  i\frac{d }{d t}\begin{pmatrix} a \\ b \end{pmatrix}=-\frac{1}{2}\begin{pmatrix} \alpha t+f(t) & v \\ v(1-\delta) & -\alpha t-f(t) \end{pmatrix}\begin{pmatrix} a \\ b \end{pmatrix}.
\end{equation}
Here, $(a,b)^{T}$ is the two-mode wave function, and $f(t)$ represents the colored noise arising from the system-bath interaction and satisfies
\begin{equation}\label{Eq-Colored Noise}
  \langle f(t)\rangle=0,\quad \langle f(t)f(t')\rangle=D^2 e^{-\gamma|t-t'|},
\end{equation}
where $D=\sqrt{\eta k_{B}T}$ represents the noise amplitude, with $k_{B}$ being the Boltzmann constant and $T$ the temperature of the heat bath. Such colored noise can be exactly described by the Ornstein-Uhlenbeck process, whose dynamics is governed by the stochastic differential equation~\cite{Crispin1986}
\begin{equation}\label{Eq-SDE}
  df(t)=-\gamma f(t)dt+\sqrt{2\gamma}DdW(t),
\end{equation}
where $dW(t)$ is the real Wiener increment satisfying $\langle dW(t) \rangle=0$ and $\langle [dW(t)]^{2} \rangle=dt$, with independent increments. 
\begin{figure}[t]
  \centering
  \includegraphics[width=\linewidth]{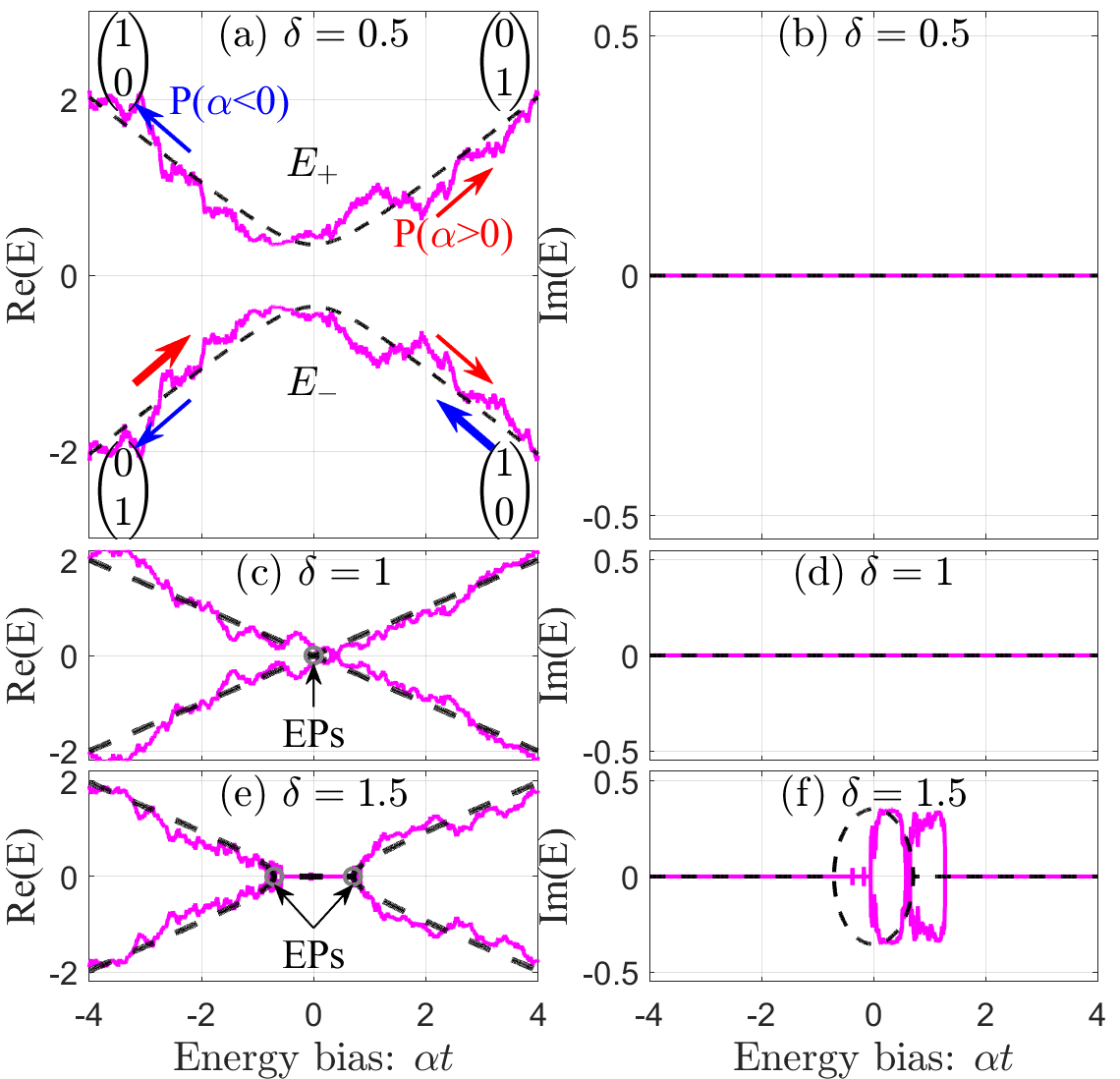}
  \caption{\label{Fig-Instantaneous_Levels}Instantaneous energy levels for various nonreciprocity parameters. The black dashed line corresponds to the noiseless case, while the magenta solid line depicts the noisy counterpart. The red and blue arrows indicate the forward and backward LZ tunneling processes, respectively.}
\end{figure}

In the absence of noise, the instantaneous eigenvalues are given by $E_{\pm}(t)=\pm\frac{1}{2}\sqrt{(\alpha t)^{2}+v^{2}(1-\delta)}$ as illustrated in Fig.~\ref{Fig-Instantaneous_Levels} and the corresponding eigenstates are $\ket{E_{\pm}(t)}=\left(\alpha t\pm\sqrt{(\alpha t)^{2}+v^{2}(1-\delta)},v(1-\delta)\right)^{T}$~\cite{Wang2022}. When $\delta>1$, the energy spectrum becomes complex as shown in Fig.~\ref{Fig-Instantaneous_Levels}(f), indicating spontaneous $\mathcal{PT}$-symmetry breaking. The eigenvalues coalesce at the EPs located at $\alpha t=\pm\sqrt{v(\delta-1)}$ (see Figs.~\ref{Fig-Instantaneous_Levels}(c)(e)). In contrast, for the noisy case, the instantaneous energy levels are obtained by directly diagonalizing the instantaneous Hamiltonian, as shown in Fig.~\ref{Fig-Instantaneous_Levels}. For $\delta>1$, the noise appreciably shifts the positions of the EPs and triggers the emergence of multiple EPs. In the case of $\delta<1$, no EPs appear even in the presence of noise, and the system retains unbroken $\mathcal{PT}$-symmetry in this regime. In the asymptotic limit $t\to\pm\infty$, the off-diagonal and noise terms become negligible compared to the diagonal level bias, and the instantaneous eigenstates reduce to $(0,1)^{T}$ and $(1,0)^{T}$ as shown in Fig.~\ref{Fig-Instantaneous_Levels}(a).

We now focus on the tunneling probability. At $t\to-\infty$, the system is initialized in the state $(a,b)^{T}=(0,1)^{T}$ for the forward sweep $\alpha>0$, and in $(1,0)^{T}$ for the backward sweep $\alpha<0$. The corresponding tunneling probabilities are defined as
\begin{equation}\label{Eq-Single Tunneling Probability}
  P(\alpha>0)=\frac{|b(t\to+\infty)|^{2}}{N(t\to+\infty)},~P(\alpha<0)=\frac{|a(t\to+\infty)|^{2}}{N(t\to+\infty)}.
\end{equation}
Here, $N=|a|^{2}+|b|^{2}$ is the total population, which is no longer conserved due to non-Hermiticity. When $\delta=0$, the system reduces to the standard LZ model, in which the tunneling probability is symmetric under sweep reversal, i.e., $P(\alpha)=P(-\alpha)=\exp\left(-\frac{\pi v^{2}}{2|\alpha|}\right)$~\cite{Landau1977}. However, for $\delta\neq0$, nonreciprocity usually leads to asymmetric tunneling probabilities such that $P(\alpha>0)\neq P(\alpha<0)$~\cite{Wang2022, Cao2024}. In the presence of noise, the tunneling probability varies among different stochastic realizations. We thus define the physical tunneling probability as the ensemble average over numerous independent realizations, that is,
\begin{equation}
  \bar{P}(\alpha)=\frac{1}{M}\sum_{i=1}^{M}P_{i}(\alpha).
\end{equation}
Here, $M$ is the total number of stochastic samples, and $P_{i}(\alpha)$ stands for the tunneling probability evaluated from the $i$-th realization according to Eq.~(\ref{Eq-Single Tunneling Probability}). In the subsequent numerical simulations, we fix the sample size to $M=2000$. The numerical convergence has been confirmed by doubling the sample size.

\section{Numerical Results}\label{Sec-Numerical Results}
In this section, we compute the ensemble-averaged tunneling probability. The effect of noise on LZ dynamics is described by two characteristic ratios that relate the noise parameters to the intrinsic scales of the unperturbed system. The ratio $\tilde{D}\equiv D/\sqrt{|\alpha|}$ quantifies the noise strength relative to the characteristic dynamical scale set by the sweep rate, and serves to delineate the weak- and strong-noise regimes. Similarly, the ratio $\tilde{\gamma}\equiv\gamma/\sqrt{|\alpha|}$ relates the noise correlation time $1/\gamma$ to the LZ transition time scale $1/\sqrt{|\alpha|}$, and further delineates the slow- and fast-noise regimes. We tune these two ratios to explore their joint effects on the tunneling probability.

To this end, we numerically solve the stochastic Schr\"odinger equation (\ref{Eq-SSE}) using the fourth-order Runge-Kutta method, during which the noise variable $f(t)$ is generated by solving Eq.~(\ref{Eq-SDE}) with the Heun scheme~\cite{Gard1988}. To specify the initial noise values for the ensemble simulations, we consider the time-dependent distribution $\Phi(f,t)$ of the noise variable, whose evolution is governed by the deterministic Fokker-Planck equation~\cite{Crispin1986}, $\frac{\partial }{\partial t}\Phi(f,t)=\gamma\frac{\partial }{\partial f}\left[f\Phi(f,t)\right]+\gamma D^{2}\frac{\partial^{2} }{\partial f^{2}}\Phi(f,t)$. Imposing the steady-state condition $\partial \Phi(f,t)/\partial t=0$ yields the stationary Gaussian distribution
\begin{equation}
  \Phi_{st}(f)=\frac{1}{\sqrt{2\pi D^{2}}}\exp\left(-\frac{f^{2}}{2D^{2}}\right).
\end{equation}
In each numerical realization, the initial value of $f$ is drawn from $\Phi_{st}(f)$ to ensure initialization at thermal equilibrium. 

\begin{figure}[t]
  \centering
  \includegraphics[width=\linewidth]{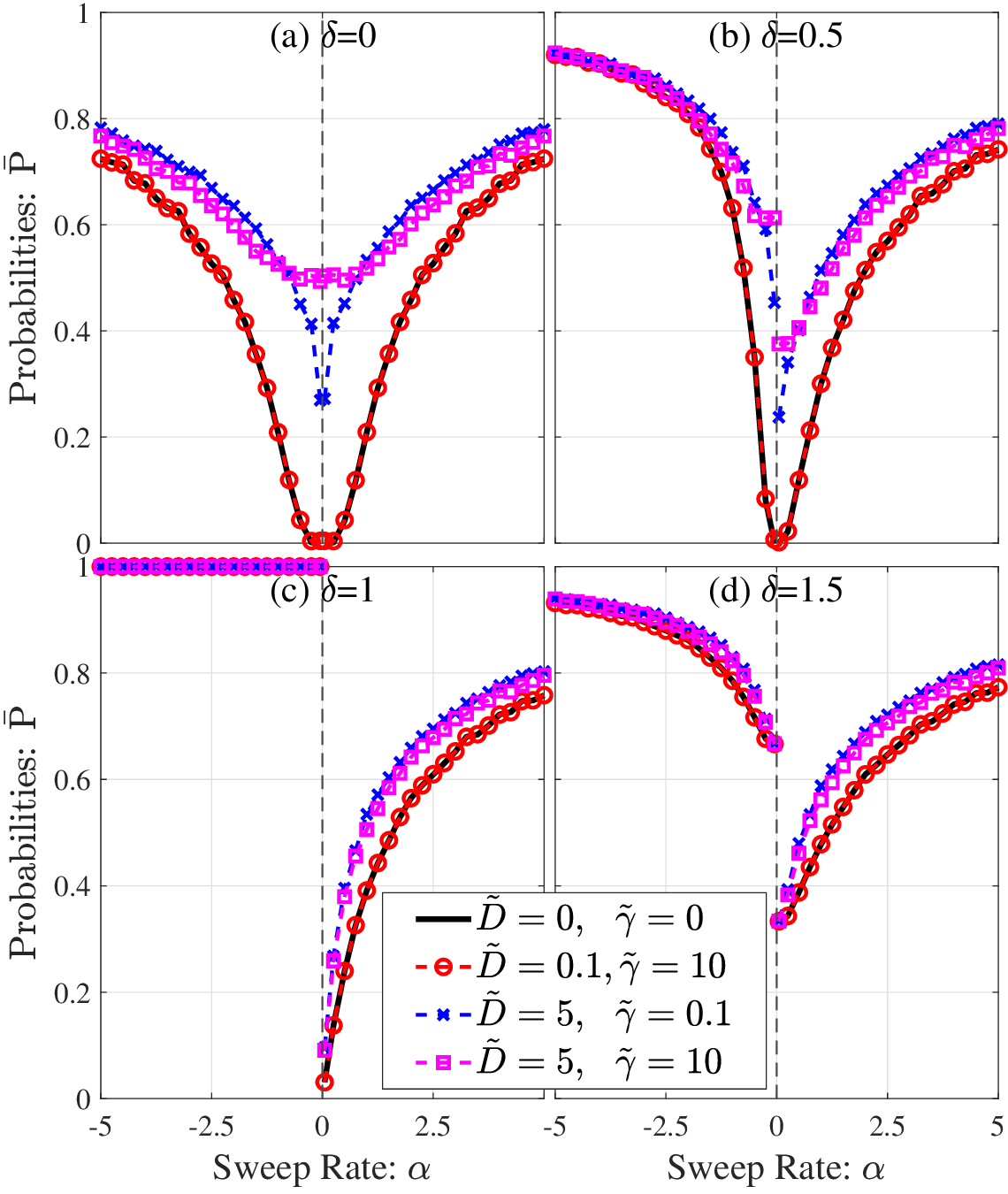}
  \caption{\label{Fig-Probabilities_with_Noise}Sweep-rate-dependent tunneling probabilities with various nonreciprocity parameters $\delta$ for different amplitudes and correlation times of colored noise.}
\end{figure}

In the Hermitian case $\delta=0$ illustrated in Fig.~\ref{Fig-Probabilities_with_Noise}(a), the forward and backward tunneling probabilities remain symmetric regardless of noise. The presence of noise always enhances tunneling. In the adiabatic limit $\alpha\to\pm0$, it breaks dynamical adiabaticity, thereby generating a finite tunneling probability. For a fixed noise amplitude, the tunneling probability is higher for fast noise than for slow noise near the adiabatic regime. As $|\alpha|$ increases, the probability for slow noise gradually exceeds that for fast noise. In the case of non-zero $\delta$, the tunneling probabilities exhibit asymmetry, i.e., $P(\alpha)\neq P(-\alpha)$. For $\delta<1$, as shown in Fig.~\ref{Fig-Probabilities_with_Noise}(b), reciprocity of LZ tunneling in the adiabatic limit is broken only in the presence of noise.

A dramatic change occurs at $\delta=1$, where the vanishing coupling term $-v(1-\delta)/2$ gives rise to energy level coalescence and the formation of EPs. In the absence of noise, the forward tunneling probability increases continuously with $\alpha$, while the backward sweep maintains a constant tunneling probability of unity, as illustrated in Fig.~\ref{Fig-Probabilities_with_Noise}(c). This behavior arises from the vanishing effective coupling during the backward sweep, which decouples the two energy levels and locks the system in its initial state, as also discussed in Refs.~\cite{Morales-Molina2011, Wang2022}. For the noisy case, the backward tunneling probability remains fixed at unity and exactly matches its noiseless counterpart. This insensitivity to noise indicates that such dynamic decoupling is robust against environmental fluctuations. 

As $\delta$ exceeds 1, the system enters the strongly non-Hermitian regime, marked by spontaneous $\mathcal{PT}$-symmetry breaking, where a purely imaginary energy gap emerges. This imaginary gap fundamentally modifies the topological characteristics of adiabatic evolution. Consequently, even in the adiabatic limit, the system cannot avoid traversing the imaginary energy region. As a result, the tunneling probabilities remain finite and converge to the same value in the adiabatic limit regardless of the noise, as shown in Fig.~\ref{Fig-Probabilities_with_Noise}(d).

\begin{figure}[t]
  \centering
  \includegraphics[width=\linewidth]{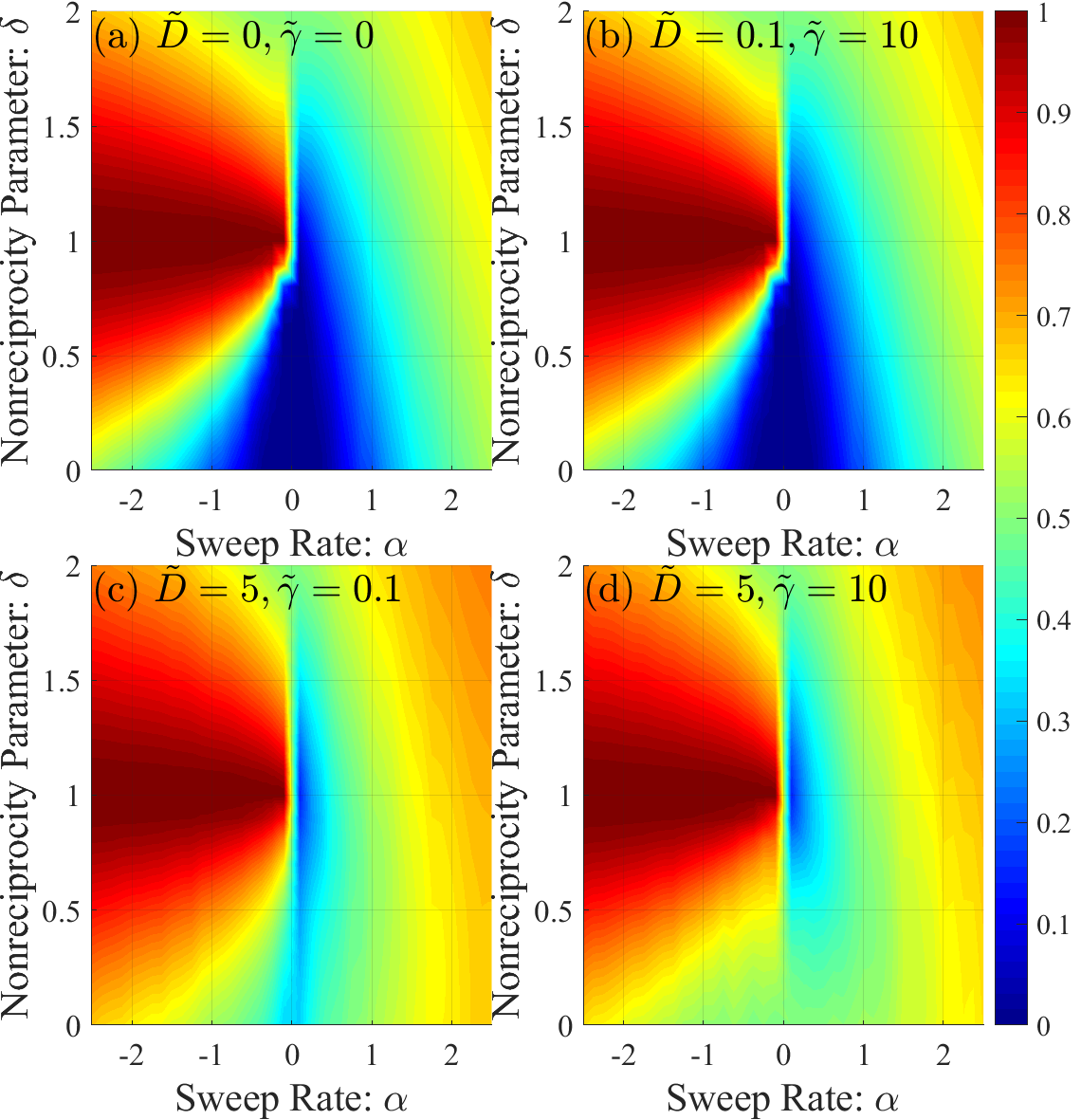}
  \caption{\label{Fig-delta_alpha_diagram}Phase diagram of the tunneling probability with respect to the sweep rate and nonreciprocity parameter for different amplitudes and correlation times of colored noise.}
\end{figure}

Furthermore, we present phase diagrams to illustrate the dependence of tunneling probability on the sweep rate and nonreciprocity parameter for several typical noise scenarios, as displayed in Fig.~\ref{Fig-delta_alpha_diagram}. In the weak-noise regime, the tunneling probability remains nearly unaffected even by fast colored noise [see Fig.~\ref{Fig-delta_alpha_diagram}(b)]. In contrast, for strong slow noise, the tunneling probability is distinctly enhanced in the parameter regime of small $|\alpha|$ and $\delta$ [see Fig.~\ref{Fig-delta_alpha_diagram}(c)]. For fixed noise amplitude, further increasing $\tilde{\gamma}$ drives the system into the fast-noise regime and makes the phase diagram become approximately symmetric with respect to the line $\delta=1$, as shown in Fig.~\ref{Fig-delta_alpha_diagram}(d).

\section{Analytical Results}\label{Sec-Analytical Results}
\subsection{Exact solutions in the absence of noise}
We first examine the tunneling probabilities in the absence of noise, where each simulation yields a unique deterministic value. For the forward sweep $\alpha>0$, the time-dependent Schr\"odinger equation can be solved analytically by decoupling its components, which reduces it to a second-order ordinary differential equation $d^{2}a/dt^{2}=\left[v^{2}(1-\delta)/4+\alpha^{2}t^{2}/4-i\alpha/2\right]a$. By introducing the dimensionless parameters $z=e^{-i\pi/4}\sqrt{\alpha}t$ and $\beta=v^{2}(1-\delta)/4\alpha$, this equation simplifies to the standard Weber equation: $d^{2}a/dz^{2}+\left(i\beta+1/2-z^{2}/4\right)a=0$. The general solution is expressed in terms of parabolic cylinder functions as $a=AD_{i\beta}(z)+BD_{-i\beta-1}(-iz)$~\cite{Whittaker2021}. Considering the initial condition of $\lim_{t_{0}\to+\infty}\psi(t=-t_{0})=(0,1)^{T}$ and using the asymptotic expressions of parabolic cylinder functions in the limit $t\to-\infty$, we can determine the parameters $A$ and $B$, that is $A=0$ and $B=\left(v/2\sqrt{\alpha}\right)e^{-\beta\pi/4}e^{i\varphi_{0}(t_{0})}$, where the phase $\varphi_{0}(t_{0})$ is defined as $\varphi_{0}(t_{0})=\pi/4+\left[v^{2}(1-\delta)/4\alpha\right]\ln(\sqrt{\alpha}t_{0})+\alpha t_{0}^{2}/4$. With the help of the asymptotic expansion of the parabolic cylinder functions, we finally obtain
\begin{subequations}
\begin{align}
  a(t_{0})=&\frac{v}{2\sqrt{\alpha}}\frac{\sqrt{2\pi}}{\Gamma\left(i\frac{v^{2}(1-\delta)}{4\alpha}+1\right)}e^{-\frac{\pi v^{2}(1-\delta)}{8\alpha}}e^{-i\frac{\pi}{4}}e^{2i\varphi_{0}(t_{0})},\\
  b(t_{0})=&e^{-\frac{\pi v^{2}(1-\delta)}{4\alpha}}.
\end{align}
\end{subequations}
Then, the total population is 
\begin{equation}
  N(+\infty)=\lim_{t_{0}\to+\infty}\left[|a(t_{0})|^{2}+|b(t_{0})|^{2}\right]=\frac{1-e^{-\frac{\pi v^{2}(1-\delta)}{2\alpha}}}{1-\delta}.
\end{equation}
The LZ tunneling probability becomes
\begin{equation}\label{Eq-ETP_forward}
  P(\alpha>0)=\lim_{t_{0}\to+\infty}\frac{|b(t_{0})|^{2}}{N(t_{0})}=\frac{(1-\delta)e^{-\frac{\pi v^{2}(1-\delta)}{2\alpha}}}{1-\delta e^{-\frac{\pi v^{2}(1-\delta)}{2\alpha}}}.
\end{equation}
The backward sweep $\alpha<0$ is treated analogously to the forward sweep, with only the initial condition replaced by $\lim_{t_{0}\to+\infty}\psi(t=-t_{0})=(1,0)^{T}$. We thus obtain the tunneling probability as
\begin{equation}\label{Eq-ETP_backward}
  P(\alpha<0)=\frac{e^{\frac{\pi v^{2}(1-\delta)}{2\alpha}}}{1-\delta+\delta e^{\frac{\pi v^{2}(1-\delta)}{2\alpha}}}.
\end{equation}
These similar results are also addressed in Refs.~\cite{Morales-Molina2011, Torosov2017}, which employ distinct initial conditions under the same sweep direction.

However, for $\delta=1$, the transformation to the Weber equation used previously is no longer valid. Since the term of $-v(1-\delta)/2$ in Hamiltonian (\ref{Eq-System Hamiltonian}) vanishes in this case, the Schr\"odinger equation can be solved analytically with initial conditions $a(-t_{0})=a_{0}$ and $b(-t_{0})=b_{0}$, which yields the exact solution $a(t_{0})=a_{0}+ivb_{0}\sqrt{\frac{\pi}{2i\alpha}}\exp\left(\frac{i\alpha t_{0}^{2}}{2}\right)\text{erf}\left(\sqrt{\frac{i\alpha}{2}}t_{0}\right),~b(t_{0})=b_{0}$. Here the symbol $\text{erf}$ denotes the error function. Using the asymptotic limit of the error function as $t_{0}\to+\infty$ and the initial states for the forward sweep $\psi(t=-t_{0})=(0,1)^{T}$ and backward sweep $\psi(t=-t_{0})=(1,0)^{T}$, the tunneling probabilities are
\begin{subequations}
\begin{align}
 P(\alpha>0)=&\lim_{t_{0}\to+\infty}\frac{|b(t_{0})|^{2}}{|a(t_{0})|^{2}+|b(t_{0})|^{2}}=\frac{2\alpha}{2\alpha+v^{2}\pi},\\
 P(\alpha<0)=&\lim_{t_{0}\to+\infty}\frac{|a(t_{0})|^{2}}{|a(t_{0})|^{2}+|b(t_{0})|^{2}}=1.
\end{align}
\end{subequations}
Ref.~\cite{Morales-Molina2011} also obtained analogous results for $\delta=1$, with the solution derived by taking the limit $\delta\to1$. These analytical results agree with those from direct numerical solution of the Schr\"odinger equation with Hamiltonian (\ref{Eq-System Hamiltonian}).

\subsection{Approximate solutions in the white noise limit}
In the framework of the Wiener-Hermite expansion~\cite{Sumi1977, Cho2011}, a bosonic operator is utilized to formulate an equivalent quantum representation of the classical colored noise process with exponential correlation defined in Eq.~(\ref{Eq-Colored Noise}). This operator takes the form $\hat{f}(t)=D\left(\hat{b}e^{-\gamma t}+\hat{b}^{\dagger}e^{\gamma t}\right)$, where $\hat{b}^{\dagger}$ and $\hat{b}$ are the bosonic creation and annihilation operators obeying the commutation relation $[\hat{b}$,~$\hat{b}^{\dagger}]=1$, and $\ket{n}$ corresponds to the associated bosonic number state. The vacuum expectation value of the time-ordered product constructed from this operator exactly reproduces the predefined noise correlation function, as expressed by $\bra{0}\hat{T}\left(\hat{f}(t)\hat{f}(t')\right)\ket{0}=D^{2}e^{-\gamma|t-t'|}$, where $\hat{T}$ is the time-ordering operator. Since $\hat{f}(t)$ is linear in the bosonic operators, all higher-order time-ordered products conform to Wick's theorem. The resulting moment structure is entirely identical to that of the classical colored noise. Consequently, the statistical average of any functional of $f(t)$ can be replaced by the time-ordered vacuum expectation value of the corresponding operator-valued functional of $\hat{f}(t)$. 

The Wiener-Hermite expansion method has been applied to study the colored noise effect on standard LZ tunneling~\cite{Kayanuma1984b, Kayanuma1985}. Here, we extend to study the nonreciprocal LZ model. For our model, we introduce a new effective Hamiltonian of the form
\begin{equation}
  \tilde{H}(t)=H_{S}-\frac{1}{2}\hat{f}\sigma_{z},
\end{equation}
where $H_{S}$ denotes the original system Hamiltonian. The corresponding density operator $\tilde{\rho}$ obeys the non-Hermitian Liouville equation $id\tilde{\rho}/dt=\tilde{H}\tilde{\rho}-\tilde{\rho}\tilde{H}^{\dagger}$. Applying the time-ordering operator $\hat{T}$ to both sides and using the fact that the time derivative commutes with time ordering yields
\begin{equation}
  i\frac{d }{d t}(\hat{T}\tilde{\rho})=\hat{T}(\tilde{H}\tilde{\rho}-\tilde{\rho}\tilde{H}^{\dagger}).
\end{equation}
We project the above equation onto the bosonic number states and take the vacuum expectation value. Then, we define the reduced matrix elements as $\chi_{i,j}(n,t)=\bra{n}(\hat{T}\tilde{\rho})_{i,j}\ket{0}$, where $i,j$ label the system degrees of freedom. Such projection removes the bosonic degrees of freedom and yields a set of deterministic differential equations for $\chi_{i,j}(n,t)$:
\begin{widetext}
\begin{subequations}
\begin{align}
  i\dot{\chi}_{1,1}(n,t)=&-\frac{v}{2}[\chi_{2,1}(n)-\chi_{1,2}(n)],\\
  i\dot{\chi}_{1,2}(n,t)=&-\alpha t\chi_{1,2}(n)-\frac{1}{2}v\chi_{2,2}(n)+\frac{1}{2}v(1-\delta)\chi_{1,1}(n)-D\left(e^{-\gamma t}\sqrt{n+1}\chi_{1,2}(n+1)+e^{\gamma t}\sqrt{n}\chi_{1,2}(n-1)\right),\\
  i\dot{\chi}_{2,1}(n,t)=&\alpha t\chi_{1,2}(n)-\frac{1}{2}v(1-\delta)\chi_{1,1}(n)+\frac{1}{2}v\chi_{2,2}(n)+D\left(e^{-\gamma t}\sqrt{n+1}\chi_{2,1}(n+1)+e^{\gamma t}\sqrt{n}\chi_{2,1}(n-1)\right),\\
  i\dot{\chi}_{2,2}(n,t)=&-\frac{1}{2}v(1-\delta)[\chi_{1,2}(n)-\chi_{2,1}(n)].
\end{align}
\end{subequations}
\end{widetext}
Using the transformation
\begin{subequations}\label{Eq-Transformation}
\begin{align}
    p(n,t)=&e^{-n \gamma t}\left(\chi_{1,1}(n,t)-\chi_{2,2}(n,t)\right), \\
    q(n,t)=&e^{-n \gamma t}\left(\chi_{1,2}(n,t)+\chi_{2,1}(n,t)\right), \\
    r(n,t)=&-i e^{-n \gamma t}\left(\chi_{1,2}(n,t)-\chi_{2,1}(n,t)\right), \\
    s(n,t)=&e^{-n\gamma t}(\chi_{1,1}(n,t)+\chi_{2,2}(n,t)),
\end{align}
\end{subequations}
the above equations can be rewritten as
\begin{widetext}
\begin{subequations}\label{Eq-WHE}
\begin{align}
  \dot{p}(n,t)=&-n\gamma p(n,t)+\frac{1}{2}v(2-\delta) r(n,t), \\
  \dot{q}(n,t)=&-n\gamma q(n,t)-\alpha tr(n,t)-D(\sqrt{n+1}r(n+1,t)+\sqrt{n}r(n-1,t)), \\
  \dot{r}(n,t)=&-n\gamma r(n,t)+\alpha tq(n,t)+\frac{1}{2}v\delta s(n,t)+\frac{1}{2}v(\delta-2)p(n,t)+D(\sqrt{n+1}q(n+1,t)+\sqrt{n}q(n-1,t)),\\
  \dot{s}(n,t)=&-n\gamma s(n,t)+\frac{1}{2}v\delta r(n,t).
\end{align}
\end{subequations}
\end{widetext}
When $\delta=0$, the above equations reduce to those in Ref.~\cite{Kayanuma1984b}. The total population of the system is governed by $s(0,t)$. For $\delta\neq 0$, nonreciprocity breaks population conservation, and the rate of population change is proportional to both $\delta$ and the coherence term $r(0,t)$. 

In the white-noise limit $\gamma\gg1$, all excited bosonic states ($n\ge1$) relax on a timescale $\sim1/\gamma$, which is far shorter than the characteristic timescale of $n=0$ state. This rapid relaxation enables us to adopt the steady-state approximation for bosonic degrees of freedom. We thus close the hierarchy in Eqs.~(\ref{Eq-WHE}) by retaining only the equations associated with the $n=0$ subspace. For brevity, we omit the index $n=0$ hereafter, namely $p(t)\equiv p(0,t)$, $q(t)\equiv q(0,t)$ and $r(t)\equiv r(0,t)$. The resulting coupled equations are presented below
\begin{subequations}\label{Eq-Zero Subspace Equations}
\begin{align}
  \dot{p}(t)=&\frac{1}{2}v(2-\delta)r(t),\\
  \dot{q}(t)=&-\alpha tr(t)-\Gamma q(t),\\
  \dot{r}(t)=&\alpha tq(t)+\frac{1}{2}v\delta s(t)+\frac{1}{2}v(\delta-2)p(t)-\Gamma r(t),\\
  \dot{s}(t)=&\frac{1}{2}v\delta r(t),
\end{align}
\end{subequations}
where $\Gamma=D^{2}/\gamma$ is the effective decoherence rate. As define in Eq.~(\ref{Eq-Transformation}), $q$ and $r$ correspond to the transverse Bloch components $\langle \sigma_{x} \rangle$ and $\langle \sigma_{y} \rangle$ on the Bloch sphere, and their magnitudes characterize the coherence of the quantum superposition~\cite{Scully1997}. The damping terms $-\Gamma q$ and $-\Gamma r$ cause these averaged quantities to decay exponentially, which drives the off-diagonal matrix elements toward zero and eventually reduces the quantum state to a classical mixture.

We now specify the initial conditions for the two sweep directions as follows. For the forward sweep, the system is prepared in state $(a,b)^{T}=(0,1)^{T}$ with the bosonic mode in its vacuum at $t=-\infty$, corresponding to $q(-\infty)=r(-\infty)=0,~p(-\infty)=-1$, and $s(-\infty)=1$. The tunneling probability is given by $P(\alpha>0)=\lim_{t\to\infty}\frac{s(t)-p(t)}{2s(t)}$. Similarly, for the backward sweep, all initial conditions for the bosonic mode remain unchanged, while the system is instead prepared in the state $(a,b)^{T}=(0,1)^{T}$. This configuration yields $q(-\infty)=r(-\infty)=0,~p(-\infty)=1$, and $s(-\infty)=1$. The corresponding tunneling probability reads $P(\alpha<0)=\lim_{t\to\infty}\frac{s(t)+p(t)}{2s(t)}$.

In the regime $\Gamma\gg1$, the fast variables $q$ and $r$ relax exponentially toward their instantaneous steady states on a timescale $\sim1/\Gamma$, whose values are governed by the slow variables $p$ and $s$. Using the steady-state approximation, we obtain the zeroth-order solutions for the slow variables $p$ and $s$. Substituting these solutions back into Eqs.~(\ref{Eq-Zero Subspace Equations}b) and (\ref{Eq-Zero Subspace Equations}c) yields the leading correction to the fast variables and refines the expression for $r$. With this corrected $r$ substituted into Eqs.~(\ref{Eq-Zero Subspace Equations}a) and (\ref{Eq-Zero Subspace Equations}d), we obtain the final state, accurate to $\mathcal{O}(\Gamma^{-2})$. A detailed derivation is provided in Appendix~\ref{Appendix-2}. The forward tunneling probability thus takes the form
\begin{widetext}
\begin{equation}\label{Eq-Approximate Solution1}
    P(\alpha>0)=\frac{(\delta-1)\left(e^{-\frac{\pi v^{2}(1-\delta)}{\alpha}}+1\right)}{\delta e^{-\frac{\pi v^{2}(1-\delta)}{\alpha}}+\delta-2}-\frac{2v^{2}(\delta-1)^{2}\left(e^{-\frac{\pi v^{2}(1-\delta)}{\alpha}}-1\right)\left(\frac{v^{4}(\delta-1)^2}{\alpha^{2}}-8\right)}{\left(\delta e^{-\frac{\pi v^{2}(1-\delta)}{\alpha}}+\delta-2\right)^{2}\left(\frac{v^{4}(\delta-1)^2}{\alpha^{2}}+4\right)\left(\frac{v^{4}(\delta-1)^2}{\alpha^{2}}+16\right)} \frac{1}{\Gamma^{2}}+\mathcal{O}\left(\frac{1}{\Gamma^{4}}\right),
\end{equation}
\end{widetext}
and the backward tunneling probability reads
\begin{widetext}
\begin{equation}\label{Eq-Approximate Solution2}
    P(\alpha<0)=\frac{e^{\frac{\pi v^{2}(1-\delta)}{\alpha}}+1}{2-\delta+\delta e^{\frac{\pi v^{2}(1-\delta)}{\alpha}}}-\frac{2v^{2}(\delta-1)^{2}\left(e^{\frac{\pi v^{2}(1-\delta)}{\alpha}}-1\right)\left(\frac{v^{4}(\delta-1)^2}{\alpha^{2}}-8\right)}{\left(2-\delta+\delta e^{\frac{\pi v^{2}(1-\delta)}{\alpha}}\right)^{2}\left(\frac{v^{4}(\delta-1)^2}{\alpha^{2}}+4\right)\left(\frac{v^{4}(\delta-1)^2}{\alpha^{2}}+16\right)} \frac{1}{\Gamma^{2}}+\mathcal{O}\left(\frac{1}{\Gamma^{4}}\right).
\end{equation}
\end{widetext}

\subsection{Some discussions}
\begin{figure}[t]
  \centering
  \includegraphics[width=\linewidth]{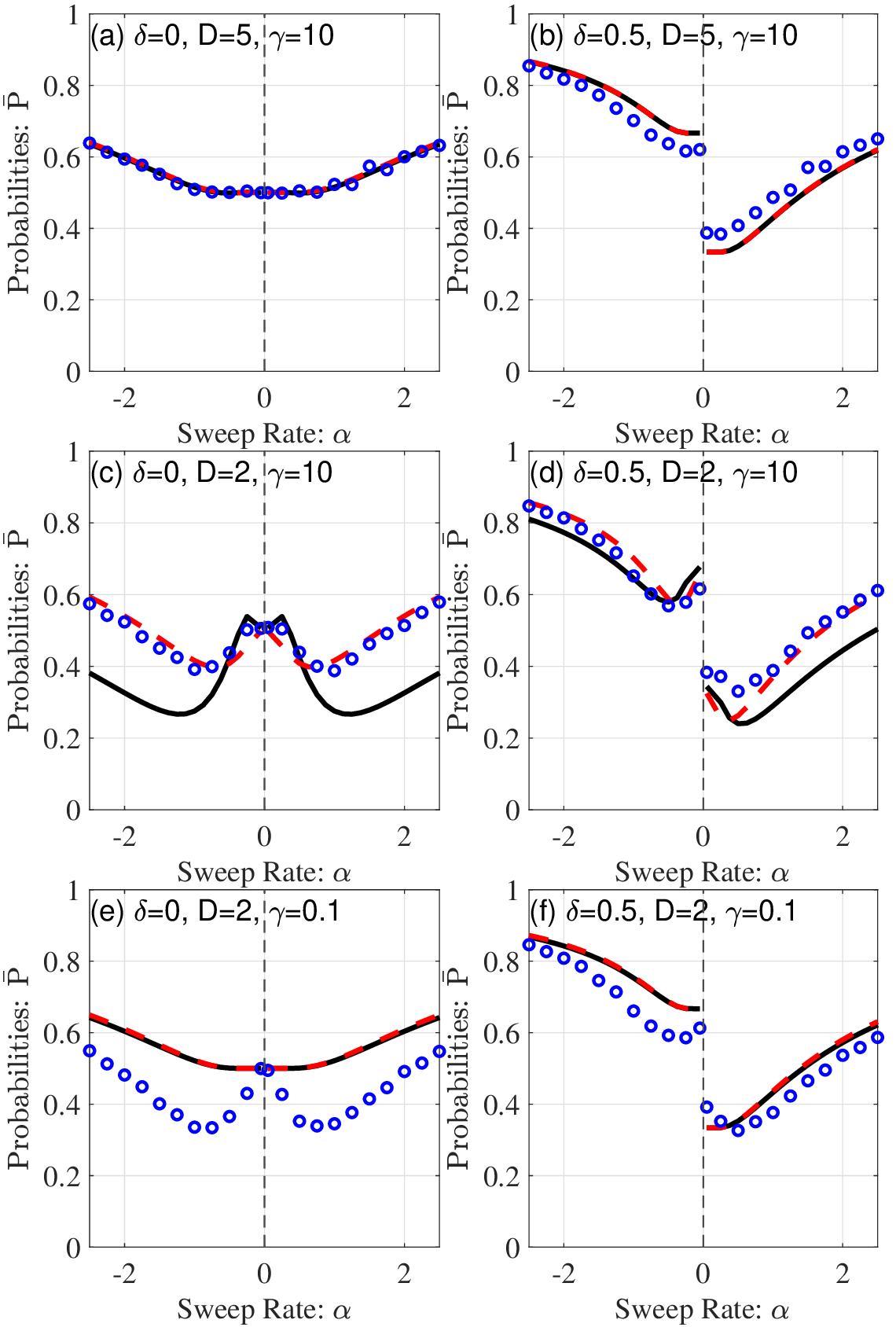}
  \caption{\label{Fig-Comparison1}Comparison of three types of solutions as functions of sweep rate: approximate solutions (black solid lines) given by Eqs.~(\ref{Eq-Approximate Solution1}), (\ref{Eq-Approximate Solution2}), numerical solutions of the $n=0$ subspace equation (red dashed lines) from Eq.~(\ref{Eq-Zero Subspace Equations}), and direct numerical solutions of the stochastic Schr\"odinger equation (blue circles) from Eq.~(\ref{Eq-SSE}). The corresponding values of $\Gamma$: (a,b) $\Gamma=2.5$; (c,d) $\Gamma=0.4$; (e,f) $\Gamma=40$.}
\end{figure}

\begin{figure}[t]
  \centering
  \includegraphics[width=\linewidth]{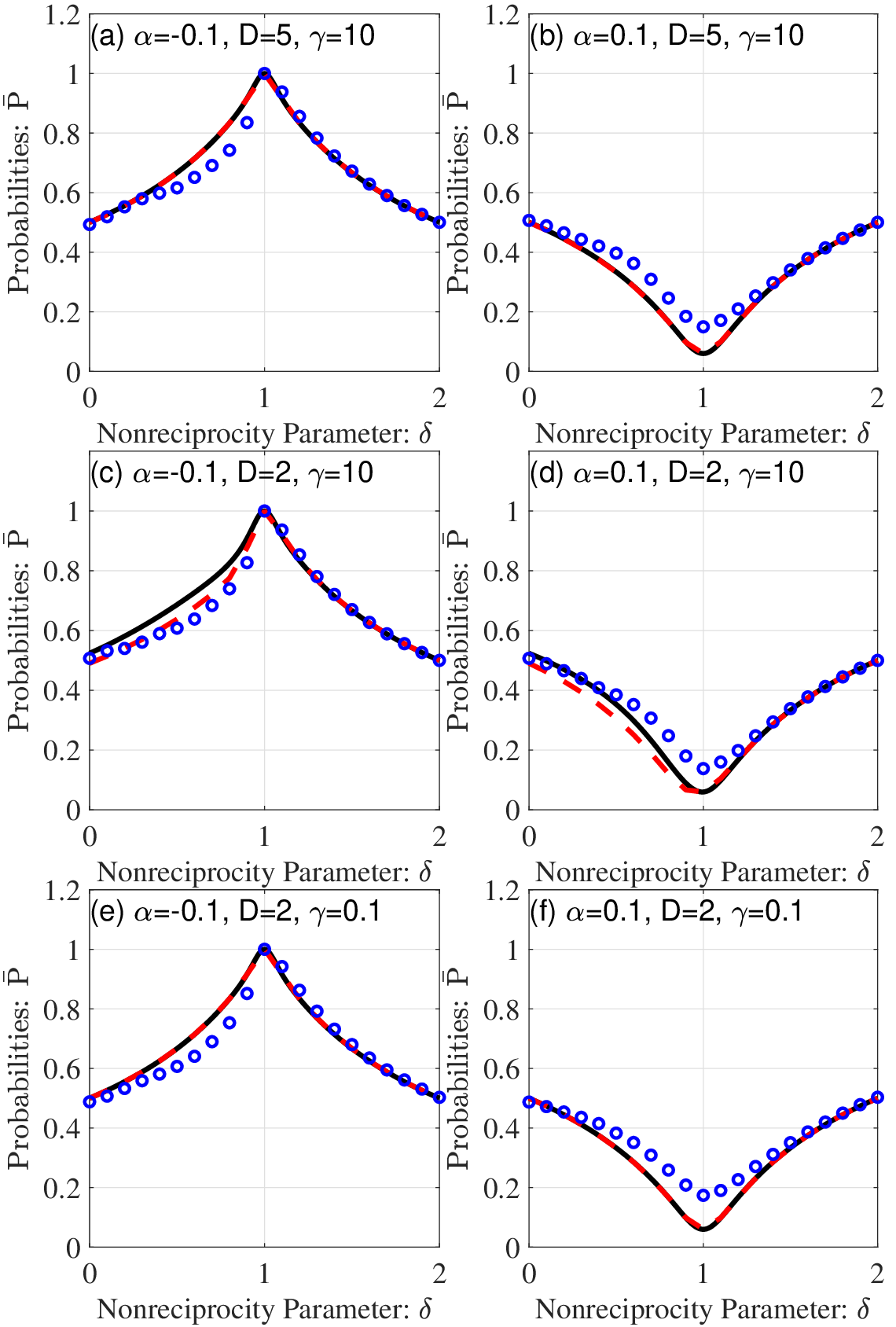}
  \caption{\label{Fig-Comparison2}Comparison of three types of solutions in the regime of nearly adiabatic sweep: approximate solutions (black solid lines) given by Eqs.~(\ref{Eq-Approximate Solution1}), (\ref{Eq-Approximate Solution2}), numerical solutions of the $n=0$ subspace equation (red dashed lines) from Eq.~(\ref{Eq-Zero Subspace Equations}), and direct numerical solutions of the stochastic Schr\"odinger equation (blue circles) from Eq.~(\ref{Eq-SSE}). The corresponding values of $\Gamma$: (a,b) $\Gamma=2.5$; (c,d) $\Gamma=0.4$; (e,f) $\Gamma=40$.}
\end{figure}

Figure~\ref{Fig-Comparison1} compares the approximate solutions in Eqs.~(\ref{Eq-Approximate Solution1}) and (\ref{Eq-Approximate Solution2}), numerical results from the $n=0$ subspace equations (\ref{Eq-Zero Subspace Equations}), and direct numerical solutions of the stochastic Schr\"odinger equation (\ref{Eq-SSE}). As displayed in Figs.~\ref{Fig-Comparison1}(a) and (b), the three sets of results agree well for sufficiently large $\gamma$ and $\Gamma$. Keeping $\gamma$ fixed, we vary parameter $D$ to violate the condition $\Gamma\gg1$, as seen in Figs.~\ref{Fig-Comparison1}(c) and (d). In this scenario, the approximate solutions lose validity, whereas numerical solutions from the $n=0$ subspace equations still agree with direct numerical results. Furthermore, in Figs.~\ref{Fig-Comparison1}(e) and (f), we fix $D$ and decrease $\gamma$ to recover the condition $\Gamma\gg1$. The approximate solutions are then in good agreement with numerical solutions of the $n=0$ subspace equations. Nevertheless, the decrease in $\gamma$ makes these $n=0$ subspace numerical solutions unreliable, which consequently diverge fully from direct numerical results. Adopting the same noise parameters as Fig.~\ref{Fig-Comparison1}, Fig.~\ref{Fig-Comparison2} compares the three solutions under nearly adiabatic sweep. All solutions are almost consistent in the near-adiabatic regime.

Comparing Eqs.~(\ref{Eq-ETP_forward}), (\ref{Eq-ETP_backward}) with (\ref{Eq-Approximate Solution1}), (\ref{Eq-Approximate Solution2}), we can prove that the noise enhances the tunneling probability in the limit $\gamma\gg1$ and $\Gamma\to\infty$, which is consistent with our previous numerical results shown in Fig.~\ref{Fig-Probabilities_with_Noise}. Such enhancement originates from heat-bath-induced thermal fluctuations, which drive thermal excitation of the system across the energy gap~\cite{Christie2024}. This incoherent excitation process directly interplays with coherent LZ tunneling, thus raising the overall tunneling probability. 

When $\Gamma\to\infty$ and $\delta=0$, Eqs.~(\ref{Eq-Approximate Solution1}) and (\ref{Eq-Approximate Solution2}) reduce to $P(\alpha>0)=P(\alpha<0)=\frac{1}{2}\left[1+\exp\left(-\frac{\pi v^{2}}{|\alpha|}\right)\right]$, which is also obtained by the formal perturbation expansion with respect to the off-diagonal coupling~\cite{Kayanuma1984a}. The noise-modulated tunneling probability can decompose into two physically distinct components. The constant term $1/2$ describes the classical statistical behavior of the system. In the limit $\Gamma\to\infty$, fast noise drives $q$ and $r$, which are constructed from the off-diagonal elements of the density matrix, to rapidly relax toward the instantaneous steady state determined by the population difference $p$, thereby destroying interlevel quantum coherence. This decoherence effect reduces Eq.~(\ref{Eq-Zero Subspace Equations}a) to the effective relaxation dynamics $\dot{p}=-\frac{v^{2}\Gamma}{\alpha^{2}t^{2}+\Gamma^{2}}p$, which drives $p$ to vanish at $\alpha=0$ and gives rise to the classical equal-population distribution. However, for $\alpha\neq0$, the decoherence-induced effective relaxation of the population difference diminishes progressively as the level bias $|\alpha t|$ increases, so that its time-integrated contribution over the entire sweep remains finite. Consequently, the population difference fails to relax to zero, and a correction term $\frac{1}{2}e^{-\frac{\pi v^{2}}{|\alpha|}}$ is retained. Compared with the standard LZ tunneling probability $\exp\left(-\frac{\pi v^{2}}{2|\alpha|}\right)$, this correction term is equivalent to half the standard LZ probability corresponding to an effective sweep rate halved by noise.

When $\delta>1$ in the adiabatic limit $\alpha\to\pm0$, Eqs.~(\ref{Eq-Approximate Solution1}) and (\ref{Eq-Approximate Solution2}) reduce to $P(\alpha\to0^{+})=(\delta-1)/\delta$ and $P(\alpha\to0^{-})=1/\delta$, respectively. Physically, the purely imaginary energy gap between EPs forms a sink and a source in the phase space of relative phase and population difference~\cite{Fu2024}. Once the quantum state falls into this sink, the population difference remains unchanged~\cite{Wang2023} and is determined by the eigenstate of EPs in the form of $\left(\pm\frac{1}{\sqrt{\delta}},\frac{\sqrt{\delta-1}}{\sqrt{\delta}}\right)^{T}$. Since the sink is global, meaning that all trajectories in phase space converge to it, the system is inherently insensitive to noise. This is verified by comparing adiabatic tunneling probabilities in noisy and noiseless scenarios, which are found to be identical, as shown in Fig.~\ref{Fig-Probabilities_with_Noise}. The same adiabatic tunneling probabilities were also obtained in Ref.~\cite{Wang2022}.

When $\delta<1$ in the adiabatic limit $\alpha\to\pm0$, Eqs.~(\ref{Eq-Approximate Solution1}) and (\ref{Eq-Approximate Solution2}) reduce to $P(\alpha\to0^{+})=(\delta-1)/(\delta-2)$ and $P(\alpha\to0^{-})=1/(2-\delta)$, respectively. By contrast, the tunneling probability vanishes without noise. This clearly demonstrates that noise breaks the reciprocity of LZ tunneling within the adiabatic limit. In the presence of noise, the forward tunneling probability decreases gradually with growing $\delta$ and approaches zero in the limit $\delta\to1$, while the backward one increases monotonically and converges to unity.

\section{Conclusion}\label{Sec-Conclusion}
In conclusion, we have investigated nonreciprocal LZ tunneling under heat-bath-induced colored noise, and analyzed how noise modulates instantaneous energy levels and tunneling probabilities. The results reveal that noise generally enhances the tunneling probability, whereas such probability becomes insensitive to noise in the adiabatic limit when the nonreciprocal parameters are sufficiently large. Moreover, noise breaks the reciprocity of LZ tunneling within the adiabatic regime. By employing Weber functions and the Wiener-Hermite expansion method, we derive the exact tunneling probability for the noiseless case and obtain an approximate solution in the white-noise limit. 

Experimentally, a double-well potential with atoms injected into one well and removed from another well can effectively realize a $\mathcal{PT}$-symmetric potential within the mean-field model~\cite{Konotop2016}. Tilting this double-well potential may then give rise to the nonreciprocal LZ dynamics. Our results deepen the physical understanding of the joint effects of non-Hermiticity and noise on quantum tunneling dynamics. This work also offers useful guidelines for quantum control and quantum simulation, and may stimulate further experimental studies along this direction.

\begin{acknowledgments}
This work was supported by the Science Challenge Project (Grant No.~TZ2025017) and National Natural Science Foundation of China (Grant No.~U2330401).
\end{acknowledgments}

\appendix
\section{Derivation of the Stochastic Schr\"odinger Equation Eq.~(\ref{Eq-SSE})}\label{Appendix-1}
In this appendix, we provide a detailed derivation of the stochastic Schr\"odinger equation (\ref{Eq-SSE}) presented in the main text. For notational convenience, we define the system operator $S\equiv-\frac{1}{2}\sigma_{z}$. The Heisenberg equations of motion for the bath oscillators take the form $\dot{q}_{k}=p_{k},~\dot{p}_{k}=-\omega_{k}^{2}q_{k}-\eta_{k}S$. Eliminating $p_{k}$ yields $\ddot{q}_{k}+\omega_{k}^{2}q_{k}=-\eta_{k}S$, whose general solution reads
\begin{equation}\label{Eq-Bath Solution}
  q_{k}(t)=q_{k}^{(h)}(t)-\eta_{k}\int_{0}^{t}dt'\frac{\sin[\omega_{k}(t-t')]}{\omega_{k}}S(t').
\end{equation}
Here, $q_{k}^{(h)}(t)$ is the homogeneous solution
\begin{equation}
  q_{k}^{(h)}(t)=q_{k}(0)\cos\omega_{k}t+p_{k}(0)\frac{\sin(\omega_{k}t)}{\omega_{k}},
\end{equation}
which is determined solely by the initial coordinate $q_{k}(0)$ and momentum $p_{k}(0)$ of the bath oscillators. Substituting Eq.~(\ref{Eq-Bath Solution}) into the Heisenberg equation of motion for an arbitrary system operator $O$, we obtain
\begin{equation}\label{Eq-Heisenberg equation of motion}
\begin{split}
  \dot{O}(t)=&\frac{1}{i\hbar}[O(t),H_{S}]+\frac{1}{i\hbar}[O(t),S]\xi(t)\\
  &-\frac{1}{i\hbar}[O(t), S]\int_{0}^{t}dt'K(t-t')S(t'),
\end{split}
\end{equation}
where
\begin{equation}
  \xi(t)=\sum_{k}\eta_{k}q_{k}^{(h)}(t),\quad K(t)=\sum_{k}\frac{\eta_{k}^{2}}{\omega_{k}}\sin(\omega_{k}t).
\end{equation}
We assume that the system and bath are initially decoupled, and the bath at temperature $T$ is initially in thermal equilibrium, as described by the density matrix $\rho_{B}(0)=\frac{e^{-H_{B}/(k_{B}T)}}{\text{Tr}e^{-H_{B}/(k_{B}T)}}$, where $k_{B}$ is the Boltzmann constant. The correlation function of $\xi(t)$ is then given by~\cite{Ford1988}
\begin{eqnarray}
  \notag
  \langle\xi(t)\xi(t')\rangle=&\sum_{k}\frac{\eta_{k}^{2}\hbar}{2\omega_k}\bigg\{\coth\left(\frac{\hbar\omega_k}{2k_{B}T}\right)\cos[\omega_k(t-t')]\\
  &-i\sin[\omega_k(t-t')]\bigg\}.
\end{eqnarray}
One can readily verify that $\langle \xi(t)\rangle=0$. This vanishing mean value follows directly from thermal equilibrium distribution and the linear dependence of $\xi(t)$ on the bath coordinates and momenta. In the continuum limit for the bath oscillators, we introduce the spectral density $J(\omega)=\frac{\pi}{2}\sum_{k}\frac{\eta_{k}^{2}}{\omega_{k}}\delta(\omega-\omega_{k})$. The correlation function can then be expressed as a continuous integral
\begin{eqnarray}
  \notag
  \langle\xi(t)\xi(t')\rangle=&\frac{\hbar}{\pi}\int_{0}^{\infty}d\omega J(\omega)\bigg\{\coth\left(\frac{\hbar\omega}{2k_{B}T}\right)\cos[\omega(t-t')]\\
  &-i\sin[\omega(t-t')]\bigg\}.
\end{eqnarray}
In what follows, we choose the Drude spectral density~\cite{Leggett1987, Breuer2007, Weiss2021}
\begin{equation}
  J(\omega)=\eta\frac{\omega\gamma}{\omega^{2}+\gamma^{2}},
\end{equation}
where $\eta$ characterizes the effective system-bath coupling strength and $\gamma$ denotes the Drude cutoff frequency. In the classical limit $\hbar\to0$, employing the approximation $\coth\left(\frac{\hbar\omega}{2k_{B}T}\right)\approx\frac{2k_{B}T}{\hbar\omega}$, the correlation function of $\xi(t)$ reduces to
\begin{equation}
  \langle\xi(t)\xi(t')\rangle=\eta k_{B}Te^{-\gamma|t-t'|}-i\frac{\hbar\eta\gamma}{2}e^{-\gamma|t-t'|}.
\end{equation}
As $\hbar\to0$, the imaginary term vanishes and only the real exponential correlation survives. Consequently, $\xi(t)$ can be treated as classical colored noise, which we denote by $f(t)$ for notational clarity. This classical noise satisfies
\begin{equation}
  \langle f(t)\rangle=0,\quad \langle f(t)f(t')\rangle=\eta k_{B}Te^{-\gamma|t-t'|}.
\end{equation}
In the Markovian limit, the memory integral term in Eq.~(\ref{Eq-Heisenberg equation of motion}) reduces to a frequency shift, which can be eliminated by redefining the system phase~\cite{Gardiner2004}. The resulting Heisenberg equation of motion then simplifies to
\begin{equation}
  \dot{O}=\frac{1}{i\hbar}[O(t),H_{S}]+\frac{1}{i\hbar}[O(t),S]f(t).
\end{equation}
This is equivalent to directly adding a random term $f(t)S$ to the system Hamiltonian, and the effective Hamiltonian takes the form
\begin{equation}
  H_{eff}=-\frac{1}{2}\begin{pmatrix} \alpha t+f(t) & v \\ v(1-\delta) & -\alpha t-f(t) \end{pmatrix}.
\end{equation}

\section{Derivation of Tunneling Probability in the Regime $\Gamma\gg1$}\label{Appendix-2}
Here, we present a detailed derivation of the tunneling probability based on the set of equations (\ref{Eq-Zero Subspace Equations}) in the $n=0$ subspace in the regime $\Gamma\gg1$. For the forward sweep, the initial conditions at $t\to-\infty$ are $q(-\infty)=r(-\infty)=0,~p(-\infty)=-1$, and $s(-\infty)=1$. We define $S(t)=\frac{v}{2}[\delta s+(\delta-2)p]$, which satisfies $S(-\infty)=v$ and gives $\dot{S}=v^{2}(\delta-1)r$. In the regime $\Gamma\gg1$, $q$ and $r$ are fast variables. We thus adopt the steady-state approximation $\dot{q}\approx0$ and $\dot{r}\approx0$, which leads to $\dot{S}^{(0)}=v^{2}(\delta-1)r^{(0)}=v^{2}(\delta-1)\frac{\Gamma S^{(0)}}{\Gamma^{2}+\alpha^{2}t^{2}}$. Hereafter, the superscript $(0)$ denotes the steady-state solution. Integrating this equation yields
\begin{equation}\label{Eq-S0}
  S^{(0)}(t)=v\exp\left[v^{2}(\delta-1)I(t)\right],
\end{equation}
where $I(t)=\frac{1}{\alpha}\left(\arctan\frac{\alpha t}{\Gamma}+\frac{\pi}{2}\right)$. Then, we substitute the steady-state solution $S^{(0)}(t)$ in to Eq.~(\ref{Eq-Zero Subspace Equations}c) to obtain the first-order correction. With $z=q+ir$, Eqs.~(\ref{Eq-Zero Subspace Equations}b) and (\ref{Eq-Zero Subspace Equations}c) yield $\dot{z}=(i\alpha t-\Gamma)z+iS^{(0)}$ and $z(-\infty)=0$, whose solution is $z(t)=ie^{i\alpha t^{2}/2-\Gamma t}\int_{\infty}^{t}e^{-i\alpha\tau^{2}/2+\Gamma\tau}S^{(0)}(\tau)d\tau$. Taking the imaginary part gives
\begin{equation}
  r(t)=\text{Re}\int_{-\infty}^{t}e^{i\alpha t^{2}/2-\Gamma t}e^{-i\alpha\tau^{2}/2+\Gamma\tau}S^{(0)}(\tau)d\tau.
\end{equation}
Setting $\xi=t-\tau$, this becomes
\begin{equation}
  r(t)=\text{Re}\int_{0}^{\infty} \exp\left(-\frac{i\alpha}{2}\xi^2 + i\alpha t \xi - \Gamma \xi\right) S^{(0)}(t-\xi) d\xi.
\end{equation}
Due to the factor $e^{-\Gamma \xi}$, the effective range of $\xi$ in the regime $\Gamma\gg1$ is $\mathcal{O}(1/\Gamma)$. We therefore expand $S^{(0)}(t-\xi)$ about $\xi=0$ as $S^{(0)}(t-\xi) = S^{(0)}(t)-\xi\dot{S}^{(0)}(t)+\frac{1}{2}\xi^2 \ddot{S}^{(0)}(t)+\mathcal{O}(\xi^3),$ and expand the oscillatory phase as $e^{-i\alpha\xi^{2}/2}=1-\frac{i\alpha}{2}\xi^{2}+\mathcal{O}(\xi^{4})$. Substituting these expansions into the expression for $r(t)$, we obtain
\begin{equation}
\begin{split}
  r(t)=\text{Re}\int_{0}^{\infty}&\bigg\{\left[S^{(0)}+\dot{S}^{(0)}\xi+\frac{1}{2}\ddot{S}^{(0)}\xi^{2}-\frac{i\alpha}{2}S^{(0)}\xi^{2}+\mathcal{O}(\xi^{3})\right]\\
  &\times \exp\left(i\alpha t\xi-\Gamma\xi\right)\bigg\}d\xi.
\end{split}
\end{equation}
With $u=\alpha t/\Gamma$, the integral evaluates to
\begin{equation}
\begin{split}
  r(u)=&\frac{S^{(0)}}{\Gamma}\frac{1}{1+u^{2}}-\frac{\dot{S}^{(0)}}{\Gamma^{2}}\frac{1-u^{2}}{(1+u^{2})^{2}}+\frac{\ddot{S}^{(0)}}{\Gamma^{3}}\frac{1-3u^{2}}{(1+u^{2})^{3}}\\
  &+\frac{\alpha S^{(0)}}{\Gamma^{3}}\frac{3u-u^{3}}{(1+u^{2})^{3}}+\mathcal{O}(\Gamma^{-4}),
\end{split}
\end{equation}
where the first term corresponds to the steady-state solution $r^{(0)}$, while the remaining three terms are denoted as $r_{2},~r_{3},~r_{4}$. From Eq.~(\ref{Eq-S0}), we have $\dot{S}^{(0)}=\frac{v^{2}(\delta-1)S^{(0)}}{\Gamma}\frac{1}{1+u^{2}}$ and $\ddot{S}^{(0)}=\frac{v^{4}(\delta-1)^{2}S^{(0)}}{\Gamma^{2}}\frac{1}{(1+u^{2})^{2}}+\mathcal{O}(\Gamma^{-3})$. With these results, we rewrite
\begin{subequations}
  \begin{align}
    r_{2}=&-\frac{v^{2}(\delta-1)S^{(0)}}{\Gamma^{3}}\frac{1-u^{2}}{(1+u^{2})^{3}},\\
    r_{3}=&\frac{v^{4}(\delta-1)^{2}S^{(0)}}{\Gamma^{5}}\frac{1-3u^{2}}{(1+u^{2})^{5}},\\
    r_{4}=&\frac{\alpha S^{(0)}}{\Gamma^{3}}\frac{3u-u^{3}}{(1+u^{2})^{3}}.
  \end{align}
\end{subequations}
It is clear that $r_3$ is much smaller than $r_2$ and $r_4$ and can thus be neglected. Therefore, only $r_2$ and $r_4$ contribute to the correction. The integral of $r$ thus can be written as
\begin{equation}
\begin{split}
  R=&\int_{-\infty}^{-\infty}r^{(0)}dt+\int_{-\infty}^{-\infty}(r_{2}+r_{4})dt+\mathcal{O}(\Gamma^{-4})\\
  \equiv& R^{(0)}+R_{corr}+\mathcal{O}(\Gamma^{-4}).
\end{split}
\end{equation}
The steady-state solution can be integrated directly to give
\begin{equation}
  R^{(0)}=\frac{v(e^{K\pi}-1)}{v^{2}(\delta-1)},
\end{equation}
where $K=\frac{v^{2}(\delta-1)}{\alpha}$. For the correction terms $R_{corr}$, we adopt the substitution $u=\tan\theta$ and integrate to obtain
\begin{equation}
  R_{corr}=\frac{(K^{2}-8)v(e^{K\pi}-1)}{(K^{2}+4)(K^{2}+16)}\frac{1}{\Gamma^{2}}.
\end{equation}
Using Eqs.~(\ref{Eq-Zero Subspace Equations}a) and (\ref{Eq-Zero Subspace Equations}d), we get
\begin{subequations}
\begin{align}
  p(\infty)=&-1+\frac{1}{2}v(2-\delta)R,\\
  s(\infty)=&1+\frac{1}{2}v\delta R.
\end{align}
\end{subequations}
The forward tunneling probability is thus given by
\begin{widetext}
\begin{equation}
    P(\alpha>0)=\frac{s(\infty)-p(\infty)}{2s(\infty)}=\frac{(\delta-1)(e^{K\pi}+1)}{\delta e^{K\pi}+\delta-2}-\frac{2v^{2}(e^{K\pi}-1)(\delta-1)^{2}}{(\delta e^{K\pi}+\delta-2)^{2}}\frac{K^{2}-8}{(K^{2}+4)(K^{2}+16)}\frac{1}{\Gamma^{2}}+\mathcal{O}\left(\frac{1}{\Gamma^{4}}\right),
\end{equation}
\end{widetext}
For the backward sweep, initial conditions and the definition of tunneling probability differ, while the solution can be derived following the same procedure.

\end{document}